Letter

# Rotation induced color confinement

Guojun Huang [a,b,*], Shile Chen [a], Yin Jiang [c], Jiaxing Zhao [d,e], Pengfei Zhuang [a]

[a] Physics Department, Tsinghua University, Beijing 100084, China
[b] School of Science and Engineering, The Chinese University of Hong Kong Shenzhen (CUHK-Shenzhen), Guangdong, 518172, China
[c] School of Physics, Beihang University, Beijing 100084, China
[d] Helmholtz Research Academy Hesse for FAIR (HFHF), GSI Helmholtz Center for Heavy Ion Physics, Campus Frankfurt, 60438 Frankfurt, Germany
[e] Institut für Theoretische Physik, Johann Wolfgang Goethe-Universität, Max-von-Laue-Straße 1, D-60438 Frankfurt am Main, Germany

ABSTRACT

The rotation effect on the QCD properties is an open question. We study the dynamic gluon mass in a dense QCD matter, the rotation is introduced by taking a covariant transformation between the flat and curved spaces. The law of causality which restricts the rotation strength of the system is carefully considered in the calculation. We find that the rotation effect is not monotonous. Overall, it behaves like an anti-screening effect, reflecting in the decreasing gluon mass, but the strength changes with the rotation. For a QCD matter with low baryon density, the screening effect in the flat space can be completely canceled by the rotation, and gluons are confined in a strongly rotating matter. When the rotation is extremely high, the matter approaches to a weakly interacting gas.

## 1. Introduction

It is widely accepted that the strongest rotation in nature can be generated in high energy nuclear collisions. It can reach $\omega \sim (9\pm1)\times 10^{21}/s$ in non-central heavy ion collisions at the Relativistic Heavy Ion Collider (RHIC) [1–3]. Since a hot and dense Quantum Chromodynamics (QCD) medium can be created in such collisions, a natural question is that how the rotation affects the phase structure of QCD. The lattice simulation of QCD with an imaginary rotation favors confinement: The phase transition temperature $T_c$ from hadron gas to quark matter increases with the rotation [4], while the effective model calculations favor deconfinement: $T_c$ drops down with the rotation [5–8]. The contradiction between lattice and models here is very different from the calculations at finite baryon density $n_B$ and/or magnetic field $B$ where $T_c$ is reduced in both lattice and models [9–11]. While the rotation effect on the QCD phase diagram is not yet clear, there are already a lot of discussions on the consequences of the rotation in heavy ion collisions. For instance, the rotation motion will polarize the spin of the finally produced hadrons via the spin-vorticity coupling [1,12,13], and one of the evidences is the observed global polarization of $\Lambda$ hyperons measured by the STAR collaboration [14,15]. The vorticity induced global and local spin alignments of vector mesons, such as $\phi$ and $K$, have also been measured in the experiments [16–18]. Moreover, such rotational collective motion in hot medium may induce anomalous transport effects as well like Chiral Vortical Effect (CVE) [19,20] and Chiral Vortical Wave (CVW) [21] which predict a baryon current or a baryonic charge quadrupole along the fluid rotation axis. In the meantime, the rotational hot medium will affect the production rate and ellipticity of lepton pairs [22,23] and quarkonium dissociation [24,25]. Besides, the rapidly rotating compact stars [26,27] provide also opportunities to investigate the cold and rotational QCD systems. To deeply understand the physics of these phenomena, it is necessary to study the rotation effect on the QCD properties beyond effective models.

In this study we calculate the gluon mass in a rotating quark matter. It is well-known that the mass of a particle controls the interaction length it propagates: Light particles propagate long-range interactions and heavy particles propagate short-range interactions. At finite temperature and baryon density, the increasing gluon mass [28] leads to a decreasing color interaction and in turn a deconfinement phase transition. Since thermodynamic functions of a system correspond to closed Feynman diagrams, calculating gluon mass in QCD requires only parton propagators. The fermion propagator in a rotating system is studied in Ref. [29]. For gluons, it is shown that [30] the gluon acquires a dynamical mass if it develops a pole at zero momentum transfer. Our recent research [31] for gluon mass indicates that there exists a resonant screening when the baryon chemical potential and the magnetic field match to each other, and the screening leads to a Shubnikov-de Haas oscillation [32] when the temperature is turned on. Since a rota-tion is similar to a magnetic field in the sense of breaking the spherical symmetry, such screening and oscillation are expected in a rotational QCD system.

* Corresponding author.
E-mail address: hgj17@tsinghua.org.cn (G. Huang).

Considering that a rotating system is equivalent to a curved space from Einstein's strong equivalence principle, we first set up the general covariant transformation between the gluon propagators in flat and curved spaces, and then perform the detailed transformation for the gluon self-energy. Taking into account the causality condition for a rotating system and using the known self-energy in flat space in the frame of QCD resummation, we calculate the gluon mass at finite baryon density and rotation. We analyze the physics of the calculation and summarize the result in the end.

## 2. Einstein's strong equivalence principle

A usual approach to take the rotation of a system into account is to go into the co-rotating frame by considering a curved metric [33]. In this way, explicit rotation-dependent terms will emerge in the Lagrangian density of the system, but the field operators still satisfy the usual periodic (for bosons) or anti-periodic (for fermions) condition along the imaginary temporal direction. We consider an alternative approach in this study. For a globally rotating system around the $z$-axis, the coordinate dependence of any field $\Phi$ in the laboratory frame should take the form

$$\Phi(t, r, \phi - \omega t, z), \tag{1}$$

and the rotation effect in the laboratory frame is purely induced by the transformation between the flat and curved spaces. The problem in this approach is at finite temperature. For a thermal system the rotation will change the usual periodic or anti-periodic condition into a twisted one [5,34]

$$\Phi(\tau, r, \phi, z) = \Phi(\tau + \beta, r, \phi + \omega\beta, z) \tag{2}$$

with $\tau = it$ and $\beta = 1/T$ in the imaginary time formalism of finite temperature field theory. To avoid this problem we focus in this study on the density and rotation effects on the QCD matter.

According to Einstein's strong equivalence principle, the gravity or curved space-time can be treated as a non-inertial system [35], and vice versa, a rotating system can be equally described as a curved or gravity-like frame by taking the covariant transformation between the two groups of coordinates in flat and curved spaces. In this way, we can obtain the rotating QCD matter's characteristics from the known properties in flat space.

Since the gluon mass is defined as the pole of the gluon propagator, we derive in the following the transformation for the gluon propagators in the flat and curved spaces. The coordinate transformation for a constant rotation $\omega$ around $z$-axis is

$$\begin{cases} t = \bar{t}, \\ x = \bar{x}\cos(\omega\bar{t}) - \bar{y}\sin(\omega\bar{t}), \\ y = \bar{x}\sin(\omega\bar{t}) + \bar{y}\cos(\omega\bar{t}), \\ z = \bar{z} \end{cases} \tag{3}$$

with coordinates $\bar{x}^\mu = (\bar{t}, \bar{x}, \bar{y}, \bar{z})$ and $x^\mu = (t, x, y, z)$ in the flat and curved spaces. From the partition function in curved space and taking its functional derivative with respect to the auxiliary field, see the details in Appendix A, the gluon propagators $D$ and $\bar{D}$ in the two spaces satisfy the relation [36],

$$D^{\mu\nu}_{ab}(x, y) = \frac{\partial x^\mu}{\partial \bar{x}^\sigma}\frac{\partial y^\nu}{\partial \bar{y}^\rho}\bar{D}^{\sigma\rho}_{ab}\left(\bar{x} - \bar{y}\right). \tag{4}$$

As for the inverse propagator $D^{-1}$, the covariant transformation should be

$$[D^{-1}]^{ab}_{\mu\nu}(x, y) = \frac{\partial \bar{x}^\sigma}{\partial x^\mu}\frac{\partial \bar{y}^\rho}{\partial y^\nu}[\bar{D}^{-1}]^{ab}_{\sigma\rho}\left(\bar{x} - \bar{y}\right). \tag{5}$$

## 3. Gluon self-energy in curved space

Since the inverse propagator is a linear summation of the free one and the self-energy,

$$[D^{-1}]^{ab}_{\mu\nu}(x, y) = [D_0^{-1}]^{ab}_{\mu\nu}(x, y) + \Pi^{ab}_{\mu\nu}(x, y), \tag{6}$$

the transformation (5) keeps valid for $D_0^{-1}$ and $\Pi$. Translating the covariant transformation from coordinate space to momentum space, the gluon self-energy in curved space can be expressed as

$$\Pi^{ab}_{\mu\nu}(q, q') = \int \frac{d^4\bar{q}}{(2\pi)^4} T^{\sigma\rho}_{\mu\nu}(\bar{q}|q, q')\bar{\Pi}^{ab}_{\sigma\rho}(\bar{q}) \tag{7}$$

with gluon momenta $q_\mu = (q_0, \boldsymbol{q}), q'_\mu = (q'_0, \boldsymbol{q}'), \bar{q} = (\bar{q}_0, \bar{\boldsymbol{q}})$ and the second-order and first-order translation tensors

$$\begin{aligned} T^{\sigma\rho}_{\mu\nu}(\bar{q}|q, q') &= h^\sigma{}_\mu(\bar{q}|q)h^\rho{}_\nu(-\bar{q}|q'), \\ h^\sigma{}_\mu(\bar{q}|q) &= \int d^4x\sqrt{-\det(g_{\alpha\beta}(x))}\frac{\partial \bar{x}^\sigma}{\partial x^\mu}e^{i(q\cdot x - \bar{q}\cdot\bar{x})}, \end{aligned} \tag{8}$$

where

$$g_{\alpha\beta}(x) = \begin{pmatrix} 1 - (x^2 + y^2)\omega^2 & -y\omega & x\omega & 0 \\ -y\omega & -1 & 0 & 0 \\ x\omega & 0 & -1 & 0 \\ 0 & 0 & 0 & -1 \end{pmatrix} \tag{9}$$

is the metric in the curved space. Note that, the rotation dependence introduced via the transformation here is complete, we did not require the rotation $\omega$ to be small. In most of the previous calculations [5,37,38] people usually took the approximation of linear $\omega$-dependence.

A unique property of a rotating system is the causality condition. For a constant rotation, to guarantee the law of causality, the size $R_{max}$ of the system is under the constraint of

$$|\omega R_{max}| < 1. \tag{10}$$

For instance for $\omega = 0.1 m_\pi$ with $m_\pi$ being the pion meson mass, there is $R_{max} \le 15$ fm which is about the maximum size of the Quark-Gluon Plasma (QGP) created in nuclear collisions at RHIC. To quantitatively involve the constraint from the causality in our treatment, we consider a cube of QCD matter with side length $L$. In this case, the angular velocity $\omega$ should be smaller than $\sqrt{2}/L$. In the following numerical calculations we will take $L = 7.5$ fm, which leads to $\omega < 0.038$ GeV. The causality changes also the continuous momentum to a discrete one

$$\begin{aligned} &\boldsymbol{q} = (2\pi/L)\boldsymbol{n}, \quad \bar{\boldsymbol{q}} = (2\pi/L)\bar{\boldsymbol{n}}, \\ &\boldsymbol{n} = (n_1, n_2, n_3), \quad \bar{\boldsymbol{n}} = (\bar{n}_1, \bar{n}_2, \bar{n}_3) \\ &n_i, \bar{n}_i \in \mathbb{Z}, \quad (i = 1, 2, 3). \end{aligned} \tag{11}$$

Using the coordinate transformation (3), the partial derivative $\partial\bar{x}^\sigma/\partial x^\mu$ in the transformation tensor $h$ can be represented as [33]

$$\frac{\partial \bar{x}^\sigma}{\partial x^\mu} = (I_{||})^\sigma{}_\mu + (I_\perp)^\sigma{}_\mu\cos(\omega\bar{t}) + iJ^\sigma{}_\mu\sin(\omega\bar{t}) - I^2_\mu I^\sigma_0\omega\bar{x} + I^1_\mu I^\sigma_0\omega\bar{y}, \tag{12}$$

where $I_{||} = diag(1, 0, 0, 1)$ and $I_\perp = diag(0, 1, 1, 0)$ are the parallel and transverse components of the 4-dimensional unit matrix $I$, and the matrix

$$J = \begin{pmatrix} 0 & 0 & 0 & 0 \\ 0 & 0 & -i & 0 \\ 0 & i & 0 & 0 \\ 0 & 0 & 0 & 0 \end{pmatrix}, \tag{13}$$

is related to the gluon polarization in the transverse plane. Similarly, the phase factor in $h$ can be expressed in terms of $\bar{x}_\mu$,

$$q\cdot x - \bar{q}\cdot\bar{x} = \bar{t}(q_0 - \bar{q}_0) + \bar{z}(q_3 - \bar{q}_3) + 2/L\left(\bar{x}\Delta_1 + \bar{y}\Delta_2\right) \tag{14}$$

with

$$\Delta_1 = \pi[n_1 \cos(\omega\bar{t}) + n_2 \sin(\omega\bar{t}) - \bar{n}_1],$$
$$\Delta_2 = \pi[n_2 \cos(\omega\bar{t}) - n_1 \sin(\omega\bar{t}) - \bar{n}_2]. \tag{15}$$

After the integration over $x_\mu$ the first-order tensor $h$ becomes

$$h^\sigma{}_\mu(\bar{q}|q) = L^3 \delta_{q_3\bar{q}_3} \int d\bar{t} e^{i\bar{t}(q_0-\bar{q}_0)} \frac{\sin\Delta_1 \sin\Delta_2}{\Delta_1\Delta_2} \times\Bigg\{ (I_{||})^\sigma{}_\mu + (I_\perp)^\sigma{}_\mu \cos(\omega\bar{t}) + iJ^\sigma{}_\mu \sin(\omega\bar{t}) -iI^\sigma{}_0 \frac{\omega L}{2}\left[\frac{I^2{}_\mu}{\Delta_1}\left(1-\frac{\Delta_1}{\tan\Delta_1}\right) - \frac{I^1{}_\mu}{\Delta_2}\left(1-\frac{\Delta_2}{\tan\Delta_2}\right)\right]\Bigg\}.$$

When the rotation vanishes, it is easy to check that the first-order tensor is reduced to a $\delta$-function to guarantee the momentum conservation, and the self-energy is exactly the one in the flat space,

$$\lim_{\omega\to0} h^\sigma{}_\mu(\bar{q}|q) = 2\pi L^3 I^\sigma{}_\mu \delta(\bar{q}-q),$$
$$\lim_{\omega\to0} \Pi^{ab}_{\mu\nu}(q,q') = 2\pi L^3 \delta(-q-q')\overline{\Pi}^{ab}_{\mu\nu}(q). \tag{16}$$

## 4. Gluon mass in dense and rotational QCD matter

In an external magnetic field or rotational field, the translation invariance is broken, and there is no more shift symmetry. The gluon propagator (6) and self-energy (7) in curved space depend separately on the coordinates $x$ and $y$ or momenta $q$ and $q'$, rather than the difference $x-y$ or $q-q'$. Since the external field affects a particle motion only in the transverse direction, one usually represents the propagator and self-energy in terms of $p_{||}$ and $p_\perp$ with $p = q - q'$, see for instance Refs. [31,39].

The gluon mass is defined as the pole of the gluon propagator. In vacuum, the propagator is a function of $p^2$, and the mass $m$ is defined by the equation $D^{-1}(p^2 = m^2) = 0$. At finite temperature, the Lorentz symmetry is broken, the propagator is a function of $p_0^2$ and $\boldsymbol{p}^2$, and the dynamical mass $m$ and screening mass $m_{scr}$ are defined by the equations $D^{-1}(p_0^2 = m^2, \boldsymbol{0}) = 0$ and $D^{-1}(0, \boldsymbol{p}^2 = -m^2_{scr})$. In a magnetic or rotational field, the dynamical mass includes a longitudinal and a transverse part $m_{||}$ and $m_\perp$. After resummation over the quark loops, gluon loops and ghost loops, the two dynamical masses are calculated in the limit of the gluon self-energy $\lim_{p_0\to0} \Pi^{ab}_{\mu\nu}(p_0, p_{||} = p_\perp = 0)$ or the equivalent limit $\lim_{q_0,q'_0\to0} \Pi^{ab}_{\mu\nu}(q_0, q'_0, \boldsymbol{q} = \boldsymbol{q}' = 0)$ [30,40],

$$\Pi^{ab}_{\mu\nu}(q_0, q'_0 \to 0, \boldsymbol{q} = \boldsymbol{q}' = 0) = -2\pi L^3 \delta^{ab}(g^\perp_{\mu\nu} m^2_\perp + g^{||}_{\mu\nu} m^2_{||}), \tag{17}$$

where the 3-momentum conservation is dropped due to $\delta_{00} = 1$, and the transverse and parallel metrics $g^\perp_{\mu\nu}$ and $g^{||}_{\mu\nu}$ are defined as $g^\perp_{\mu\nu} = diag(0,-1,-1,0)$ and $g^{||}_{\mu\nu} = diag(0,0,0,-1)$. Considering that the longitudinal mass is independent of the rotational field, we are interested in the transverse mass in this work.

At the fixed point $\boldsymbol{q}, \boldsymbol{q}' = 0$ in phase space, the time evolution of the function $\Delta_i (i = 1, 2)$ disappears automatically, $\sin\Delta_i/\Delta_i$ is reduced to $\delta_{\bar{n}_i 0}$ according to the L'Hospital's rule [41], and finally the first-order transformation tensor can be explicitly written as

$$h^\sigma{}_\mu(\bar{q}|q_0, \boldsymbol{q} = 0) = \pi L^3 \delta_{\bar{n}_3 0}\Bigg\{\Bigg[2\delta_{\bar{n}_1 0}\delta_{\bar{n}_2 0}(I_{||})^\sigma{}_\mu -iI^\sigma{}_0 \frac{L\omega}{2\pi}\left(I^2{}_\mu \delta_{\bar{n}_2 0}(1-\delta_{\bar{n}_1 0})\frac{(-1)^{\bar{n}_1}}{\bar{n}_1} -I^1{}_\mu \delta_{\bar{n}_1 0}(1-\delta_{\bar{n}_2 0})\frac{(-1)^{\bar{n}_2}}{\bar{n}_2}\right)\Bigg]\delta(\bar{q}_0 - q_0) +\delta_{\bar{n}_1 0}\delta_{\bar{n}_2 0}\sum_{s=\pm}(I_\perp + sJ)^\sigma{}_\mu \delta(\bar{q}_0 - q_0 - s\omega)\Bigg\}. \tag{18}$$

The above expression tells us that the first-order and in turn the second-order transformation tensors in the limit of $q_0 = 0$ in the curved space correspond to not only the limit of $\bar{q}_0 = 0$ in the flat space but also an extra term at $\bar{q}_0 = \pm\omega$. The extra contribution comes from the coupling between gluon polarization and the rotational field. The coupling between particle spin and rotation appears also in the transport equations for quarks [33]. The details for the computation of the second-order tensor $T^{\sigma\rho}_{\mu\nu}$ at $\boldsymbol{q} = \boldsymbol{q}' = 0$ are shown in Appendix B.

Taking the quark chemical potential $\mu_q = \mu_B/3$, one can calculate the gluon self-energy induced by all one-loop Feynman diagrams in the flat space [39,42]. Since we focus on the dense and rotational effect at zero temperature, the matter induced self-energy contains only the quark-loop contribution,

$$\overline{\Pi}^{ab}_{\mu\nu}(\bar{q}) = g^2 \int \frac{d^4p}{(2\pi)^4} \delta^{ab}\left[\gamma^\mu G(p)\gamma_\nu G(p-\bar{q})\right], \tag{19}$$

where $G$ is the free quark propagator without rotation, the quark momentum integration includes a three-momentum integration and a Matsubara frequency summation. In order to include some non-perturbative effect we take the usually used resummation over quark loops on a chain [40,43–49], the gluon self-energy defined in (6) and (7) is then exactly the quark loop (19), which can further be divided into a transverse and a longitudinal part [39,40],

$$\overline{\Pi}^{ab}_{\mu\nu}(\bar{q}) = \delta^{ab}\left[P^T_{\mu\nu}(\bar{q})\overline{\Pi}_T(\bar{q}) + P^L_{\mu\nu}(\bar{q})\overline{\Pi}_L(\bar{q})\right],$$
$$\overline{\Pi}_T(\bar{q}) = -\frac{g^2\mu_q^2}{12\pi^2}\left(\frac{2\bar{q}_0^2}{|\bar{\boldsymbol{q}}|^2}+1\right) + \frac{g^2}{384\pi^2|\bar{\boldsymbol{q}}|^3}\sum_{n,s=\pm} F_T(n\bar{q}_0, |\bar{\boldsymbol{q}}|, s\mu_q),$$
$$\overline{\Pi}_L(\bar{q}) = \frac{g^2\mu_q^2}{3\pi^2}\left(\frac{\bar{q}_0^2}{|\bar{\boldsymbol{q}}|^2}-1\right) - \frac{g^2}{192\pi^2|\bar{\boldsymbol{q}}|^3}\sum_{n,s=\pm} F_L(n\bar{q}_0, |\bar{\boldsymbol{q}}|, s\mu_q) \tag{20}$$

with functions

$$F_T(n\bar{q}_0, |\bar{\boldsymbol{q}}|, s\mu_q) = \left(\bar{q}_0^2 - |\bar{\boldsymbol{q}}|^2\right)\left(\bar{q}_0^2 + n\bar{q}_0|\bar{\boldsymbol{q}}| + 4|\bar{\boldsymbol{q}}|^2 + 4ns\mu_q\bar{q}_0 + 2s\mu_q|\bar{\boldsymbol{q}}| + 4\mu_q^2\right) \times B^s_n \ln\left(\frac{\left(B^s_n\right)^2}{(n\bar{q}_0 - |\bar{\boldsymbol{q}}|)^2}\right),$$
$$F_L(n\bar{q}_0, |\bar{\boldsymbol{q}}|, s\mu_q) = \left(\bar{q}_0^2 - |\bar{\boldsymbol{q}}|^2\right)\left(n\bar{q}_0 + 2|\bar{\boldsymbol{q}}| + 2s\mu_q\right)\left(B^s_n\right)^2 \ln\left(\frac{\left(B^s_n\right)^2}{\left(n\bar{q}_0 - |\bar{\boldsymbol{q}}|\right)^2}\right), \tag{21}$$

where $B^s_n$ is defined as $B^s_n = n\bar{q}_0 - |\bar{\boldsymbol{q}}| + 2s\mu_q$. After the Matsubara summation over quark frequency, one needs to perform an analytic continuation for the gluon frequency $\bar{q}_0$ from a pure imaginary number to a real number. Substituting the self-energy $\overline{\Pi}^{ab}_{\mu\nu}(\bar{q})$ in the flat space and the transformation $T^{\sigma\rho}_{\mu\nu}(\bar{q})$ shown in Appendix B into the self-energy $\Pi^{ab}_{\mu\nu}(q,q')$ in the curved space, and then considering the limit $q_0, q'_0 \to 0$ at $\boldsymbol{q} = \boldsymbol{q}' = 0$, the longitudinal and transverse gluon masses defined in (17) are extracted as

$$m^2_{||}(\mu_q) = \frac{g^2}{6\pi^2}\mu_q^2 \tag{22}$$

which is not affected by the rotation, and

$$m^2_\perp(\mu_q, \omega) = m^2_{||}(\mu_q) + \frac{g^2\omega^2}{48\pi^2}\Bigg\{\ln\left(1-\frac{4\mu_q^2}{\omega^2}\right)^2 - \frac{L^2\mu_q^2}{3} + \frac{1}{8}\sum_{s=\pm,n\neq0}\left(2-sA_n\right)\left(1+sA_n\right)^2 \ln\left(1+sA_n\right)^2\Bigg\} \tag{23}$$

with $A_n = L\mu_q/(\pi n)$, where the summation is over quarks and antiquarks and their transverse momentum. In the limit of $\omega \to 0$, we come back to the well-known global gluon mass $m_\perp(\mu_q) = m_{||}(\mu_q)$. Note that in the calculation of the self-energy $\overline{\Pi}$ and mass square $m^2$ we did not consider the summation over the number of colors $N_c$ and the number of flavors $N_f$.

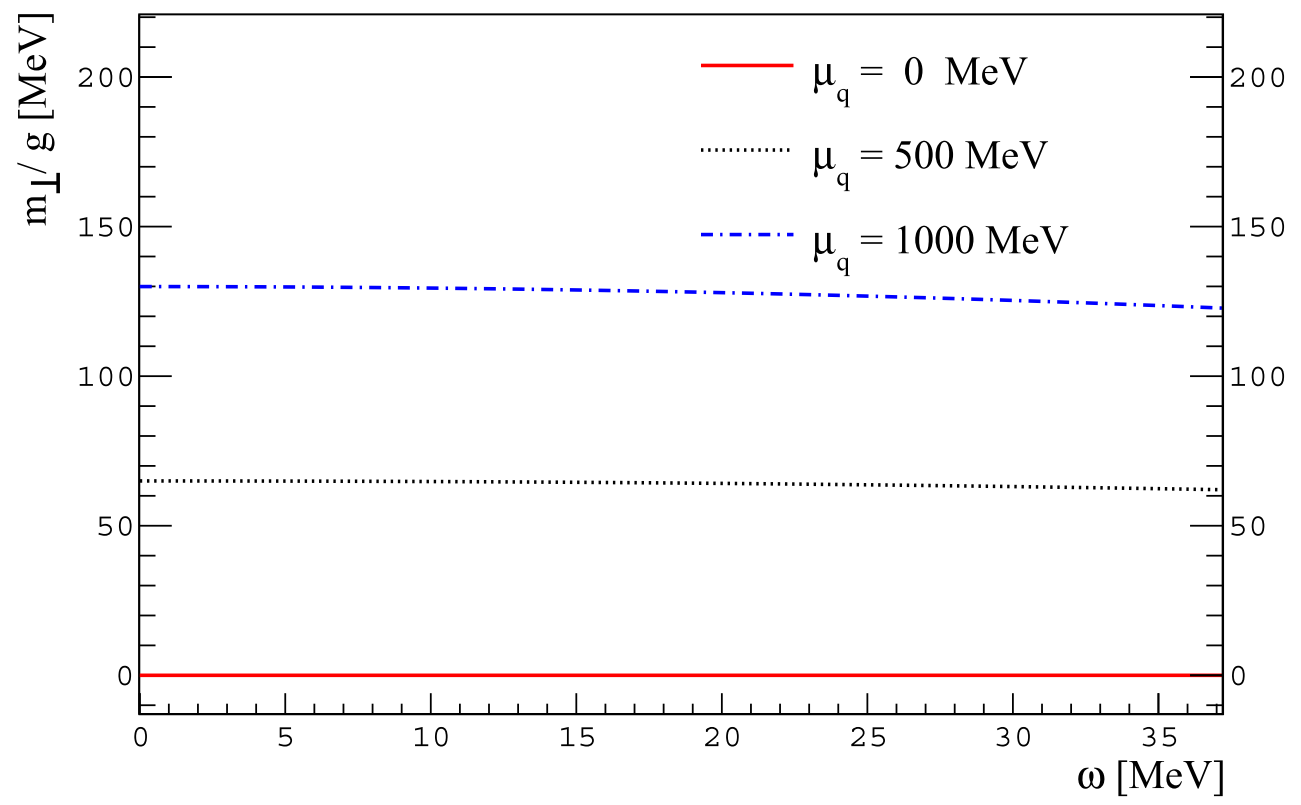

**Fig. 1.** Scaled transverse mass $m_\perp/g$ as a function of rotation at $\mu_q = 0$ (solid line), 500 (dotted line) and 1000 (dot-dashed line) MeV.

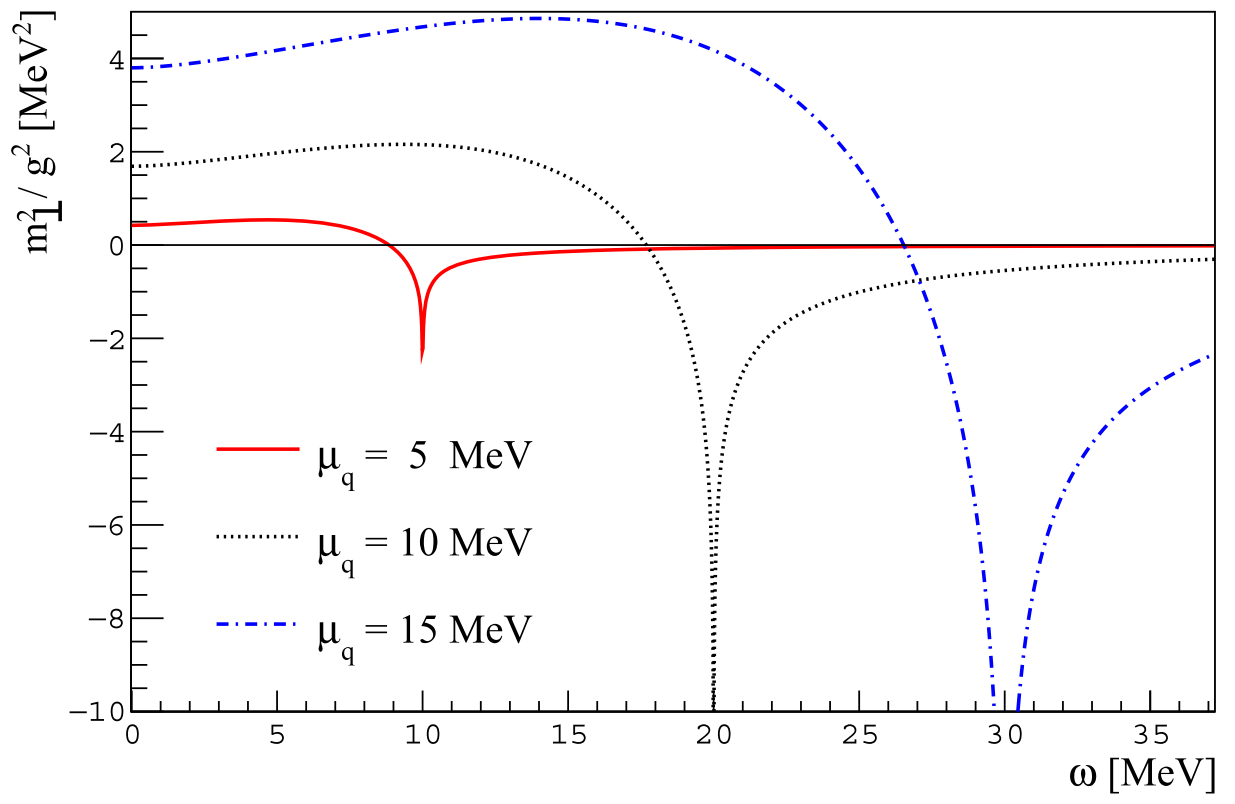

**Fig. 2.** Scaled transverse mass square $m_\perp^2/g^2$ as a function of rotation at $\mu_q = 5$ (solid line), 10 (dotted line) and 15 (dot-dashed line) MeV.

Fig. 1 shows the scaled transverse mass $m_\perp/g$ as a function of rotation $\omega$ at different quark chemical potential $\mu_q = 0$, 500 and 1000 MeV. In the vacuum without any matter ($\mu_q = 0$), there is no rotation effect, and the gluon mass keeps zero. This result is in agreement with the lattice [4] and model [5–8] calculations. As for the rotation dependence at nonzero chemical potential, gluons become massive in dense quark matter, but the rotation reduces the gluon mass slightly. This means that the density induced color screening effect is partly canceled by the rotation. Rotation is an anti-screening effect, it enhances the color interaction length and in turn favors confinement. For a finite system the global rotation must be constrained by the law od causality. For our choice of a box of quark matter with side length $L = 7.5$ fm, $\omega < 38$ MeV is much smaller in comparison with the quark chemical potentials used here. Therefore, the correction from the rotation is weak, and the maximum cancellation is less than 5%.

To see the importance of rotation, we consider now a dilute QCD matter with small chemical potential $\mu_q = 5,\ 10,\ 15$ MeV, the result is shown in Fig. 2. In this case, $\mu_q$ and $\omega$ are comparable, the rotation induced significance can be seen clearly. The rotation dependence is not monotonous. The gluon mass goes up slightly in the beginning, indicating a small enhancement of the screening, and then drops down strongly, showing a large anti-screening. The screening is completely canceled by the rotation at a critical value where gluons are again massless. After that, the anti-screening effect continuously increases, and the mass square becomes negative and reaches minus infinity at the second critical value $\omega_c$. From the analytic result (23), the divergence happens at

$$\omega_c = 2\mu_q. \tag{24}$$

After the divergence, the anti-screening effect is gradually reduced, but the mass square keeps negative. A negative $m_\perp^2$ or an imaginary mass $m_\perp$ means that gluons can no longer be considered as physical particles, the color degrees of freedom are confined. This picture of confinement is firstly introduced by Gribov [50–52] where the imaginary gluon mass is induced by the Gribov copies. The unmonotonous rotation dependence here is similar to the magnetic field effect on the chiral condensate where there is magnetic catalysis at low temperature and inverse magnetic catalysis at high temperature [10,11].

The above divergence is a general phenomenon in dense Fermionic matter, it happens in dense Quantum Electrodynamics (QED) and is called Shubnikov-de Haas oscillation [32,53]. For a dense and magnetized QCD matter, the gluon mass diverges when one of the discrete energy levels (Landau energy levels) is exactly at the Fermi surface $\sqrt{2n|qB|} = \mu_q$ [31]. For the dense and rotational QCD matter discussed here, when the spin-rotation coupling energy matches the Fermi surface $\omega/2 = \mu_q$, the gluon mass diverges. Since there can be many Landau levels ($n = 0, 1, 2, \cdots$), many divergences will emerge with increasing magnetic field. For the rotational matter, however, there is only one positive energy level $\omega/2$, there is maximum one divergence for the gluon mass at $\omega = 2\mu_q$.

## 5. Summary

We now summarize the work. The usual way to study a rotational system is in the co-rotating frame with an extra rotation term in the Lagrangian. We considered in this work an alternative way to introduce rotation in a dense QCD system in the laboratory frame, by taking a covariant transformation between the flat and curved spaces. Calculating the transformation for the gluon propagator, and taking the known propagator including the quark-loop resummation in the flat space, we obtained the gluon mass in the curved space as a function of quark chemical potential $\mu_q$ and rotation $\omega$. In the calculation we considered carefully the law of causality for a rotating system, which restricts the value of the rotation and the size of the gluon phase space. We found that the rotation effect is not monotonous. Overall, the rotation is an anti-screening effect, it favors confinement, but the strength changes with rotation. For strong enough rotation, the density induced screening effect is completely canceled, and gluons are confined with an imaginary mass, similar to the case in the Gribov approach [50–52]. When the rotation energy is exactly at the Fermi surface, $\omega/2 = \mu_q$, the anti-screening becomes infinity, leading to an infinite imaginary gluon mass. With further increasing rotation, the anti-screening effect becomes weaker and weaker.

## Declaration of competing interest

The authors declare the following financial interests/personal relationships which may be considered as potential competing interests: Guojun Huang reports financial support was provided by National Natural Science Foundation of China under Grant Nos. 12247185. If there are other authors, they declare that they have no known competing financial interests or personal relationships that could have appeared to influence the work reported in this paper.

## Acknowledgement

The work is supported by the Natural Science Foundation of China (NSFC) under Grant Nos. 12075129, 12247185, 12375131 445 (Y. J.) and the Shuimu Tsinghua Scholar Program.

## Appendix A. Covariant transformation

The action and partition function in a curved space are defined as [36]

$$S[G_a^{\ \mu}, J^a_{\ \mu}] = \int d^4x\sqrt{-\det(g_{\mu\nu}(x))} \times\left[\mathcal{L}\left(G_a^{\ \mu}(x), G_{a\ ;\nu}^{\ \mu}(x)\right)+\frac{1}{\xi}\left(G_{a\ ;\alpha}^{\ \alpha}(x)\right)^2+J^a_{\ \mu}(x)G_a^{\ \mu}(x)\right],$$
$$Z[J^a_{\ \mu}] = \int DG_a^{\ \mu}\sqrt{-\det(g_{\mu\nu}(x))}e^{iS[G_a^{\ \mu},J^a_{\ \mu}]}. \quad (A.1)$$

The gluon propagator can be obtained by the functional derivative with respect to the auxiliary field $J^a_{\ \mu}(x)$,

$$D_{ab}^{\mu\nu}(x,y) = -[\det(g_{\mu\nu}(x))\det(g_{\mu\nu}(y))]^{-1/2}\frac{1}{Z[J^a_{\ \mu}]}\frac{\delta^2 Z[J^a_{\ \mu}]}{\delta J^a_{\ \mu}(x)\delta J^b_{\ \nu}(y)}. \quad (A.2)$$

Taking into account the transformation for the auxiliary field and the definition for the $\delta$ function in the curved space,

$$J^a_{\ \sigma}(\overline{x}) = \frac{\partial x^\mu}{\partial \overline{x}^\sigma}J^a_{\ \mu}(x),$$
$$\delta^4(x-y) = \frac{1}{|\det(\partial x^\mu/\partial\overline{x}^\sigma)|}\delta^4(\overline{x}-\overline{y}), \quad (A.3)$$

there is

$$\frac{\delta}{\delta J^a_{\ \mu}(x)} \Leftrightarrow \frac{\partial x^\mu}{\partial \overline{x}^\sigma}\frac{1}{|\det(\partial x^\mu/\partial\overline{x}^\sigma)|}\frac{\delta}{\delta J^a_{\ \sigma}(\overline{x})}. \quad (A.4)$$

Considering the relations

$$|\det(\partial x^\mu/\partial\overline{x}^\sigma)| = [-\det(g_{\mu\nu}(x))]^{-1/2},$$
$$Z[J^a_{\ \mu}(x)] = Z[J^a_{\ \sigma}(\overline{x})], \quad (A.5)$$

the gluon propagator in the curved space can be represented by the one in the flat space,

$$D_{ab}^{\mu\nu}(x,y) = -\frac{\partial x^\mu}{\partial\overline{x}^\sigma}\frac{\partial y^\nu}{\partial\overline{y}^\rho}\frac{1}{Z[J^a_{\ \mu}]}\frac{\delta^2 Z[J^a_{\ \mu}]}{\delta J^a_{\ \sigma}(\overline{x})\delta J^b_{\ \rho}(\overline{y})}$$
$$= \frac{\partial x^\mu}{\partial\overline{x}^\sigma}\frac{\partial y^\nu}{\partial\overline{y}^\rho}\overline{D}_{ab}^{\sigma\rho}(\overline{x}-\overline{y}). \quad (A.6)$$

## Appendix B. Gluon mass

Using the first-order transformation tensor $h^\sigma_{\ \mu}$ shown in (18), the second-order tensor at the fixed phase-space point $\boldsymbol{q}=\boldsymbol{q}'=0$ can be expressed as

$$T_{\mu\nu}^{\sigma\rho}(\overline{q}|q_0,q_0',\boldsymbol{q}=\boldsymbol{q}'=0)$$
$$= h^\sigma_{\ \mu}(\overline{q}|q_0,\boldsymbol{q}=0)h^\rho_{\ \nu}(-\overline{q}|q_0',\boldsymbol{q}'=0)$$
$$= \pi^2L^6\delta(-q_0-q_0')\delta_{\overline{n}_30}\Big\{\delta_{\overline{n}_10}\delta_{\overline{n}_20}\Big[4(I_{||})^\sigma_{\ \mu}(I_{||})^\rho_{\ \nu}\delta(\overline{q}_0-q_0)$$
$$+\left(I_\perp+J\right)^\sigma_{\ \mu}\left(I_\perp-J\right)^\rho_{\ \nu}\delta(\overline{q}_0-q_0-\omega)$$
$$+\left(I_\perp-J\right)^\sigma_{\ \mu}\left(I_\perp+J\right)^\rho_{\ \nu}\delta(\overline{q}_0-q_0+\omega)\Big]$$
$$+\delta^\sigma_{\ 0}\delta^\rho_{\ 0}\left(\frac{L\omega}{2\pi}\right)^2\left[\frac{\delta_{\overline{n}_20}(1-\delta_{\overline{n}_10})}{\overline{n}_1^2}\delta^2_{\ \mu}\delta^2_{\ \nu}+\frac{\delta_{\overline{n}_10}(1-\delta_{\overline{n}_20})}{\overline{n}_2^2}\delta^1_{\ \mu}\delta^1_{\ \nu}\right]$$
$$\times\delta(\overline{q}_0-q_0)\Big\}+\pi^2L^6\delta_{\overline{n}_10}\delta_{\overline{n}_20}\delta_{\overline{n}_30}$$
$$\times\sum_{s=\pm}\Big\{2(I_{||})^\sigma_{\ \mu}\delta(\overline{q}_0-q_0)(I_\perp+sJ)^\rho_{\ \nu}\delta(-q_0-q_0'-s\omega)$$
$$+2(I_{||})^\rho_{\ \nu}\delta(-\overline{q}_0-q_0')(I_\perp+sJ)^\sigma_{\ \mu}\delta(-q_0'-q_0-s\omega)$$
$$+(I_\perp+sJ)^\sigma_{\ \mu}(I_\perp+sJ)^\rho_{\ \nu}\delta(\overline{q}_0-q_0-s\omega)\delta(-q_0-q_0'-2s\omega)\Big\}. \quad (B.1)$$

The terms in the first curly brace are with a common factor $\delta(-q_0-q_0')$ which goes to $\delta(0)=1$ in the limit of $q_0,q_0'\to 0$. As for the other terms in the second brace, considering the self-energy in the flat space at $\overline{\boldsymbol{q}}=0$,

$$\overline{\Pi}^{ab}_{\sigma\rho}(\overline{q}_0,\overline{\boldsymbol{q}}=0) = \delta^{ab}(g^\perp_{\sigma\rho}+g^{||}_{\sigma\rho})\left\{-\frac{g^2\mu_q^2}{6\pi^2}-\frac{g^2\overline{q}_0^2}{48\pi^2}\ln\left(1-4\mu_q^2/\overline{q}_0^2\right)^2\right\} \quad (B.2)$$

and the following summations over $\sigma$ and $\rho$ at fixed $\mu$ and $\nu$,

$$(g^\perp_{\sigma\rho}+g^{||}_{\sigma\rho})(\delta_{||})^\sigma_\mu(\delta_\perp+sJ)^\rho_\nu = 0,$$
$$(g^\perp_{\sigma\rho}+g^{||}_{\sigma\rho})(\delta_{||})^\rho_\nu(\delta_\perp+sJ)^\sigma_\mu = 0,$$
$$(g^\perp_{\sigma\rho}+g^{||}_{\sigma\rho})(\delta_\perp+sJ)^\sigma_\mu(\delta_\perp+sJ)^\rho_\nu = 0, \quad (B.3)$$

they do not contribute to the self-energy in the curved space $\Pi^{ab}_{\mu\nu}\sim T^{\sigma\rho}_{\mu\nu}\overline{\Pi}^{ab}_{\sigma\rho}$ even at $q_0,q_0'\neq 0$. Therefore, the self-energy controlling the gluon mass becomes

$$\Pi^{ab}_{\mu\nu}(q_0,q_0'\to 0,\boldsymbol{q}=\boldsymbol{q}'=0)$$
$$= \frac{\pi}{2}L^3\Big\{4(I_{||})^\sigma_{\ \mu}(I_{||})^\rho_{\ \nu}\overline{\Pi}^{ab}_{\sigma\rho}(\overline{q}_0\to 0,\overline{\boldsymbol{q}}=0)$$
$$+(I_\perp+J)^\sigma_{\ \mu}(I_\perp-J)^\rho_{\ \nu}\overline{\Pi}^{ab}_{\sigma\rho}(\overline{q}_0\to\omega,\overline{\boldsymbol{q}}=0)$$
$$+(I_\perp-J)^\sigma_{\ \mu}(I_\perp+J)^\rho_{\ \nu}\overline{\Pi}^{ab}_{\sigma\rho}(\overline{q}_0\to-\omega,\overline{\boldsymbol{q}}=0)$$
$$+\left(\frac{L\omega}{2\pi}\right)^2\Big[\sum_{\overline{n}_1\neq 0}\frac{\delta^2_{\ \mu}\delta^2_{\ \nu}}{\overline{n}_1^2}\overline{\Pi}^{ab}_{00}(\overline{q}_0\to 0,\overline{\boldsymbol{q}}=2\pi(\overline{n}_1,0,0)/L)$$
$$+\sum_{\overline{n}_2\neq 0}\frac{\delta^1_{\ \mu}\delta^1_{\ \nu}}{\overline{n}_2^2}\overline{\Pi}^{ab}_{00}(\overline{q}_0\to 0,\overline{\boldsymbol{q}}=2\pi(0,\overline{n}_2,0)/L)\Big]\Big\}. \quad (B.4)$$

Taking into account the symmetry in the flat space,

$$\overline{\Pi}^{ab}_{\sigma\rho}(\overline{q}_0,\overline{\boldsymbol{q}}=0) = \overline{\Pi}^{ab}_{\sigma\rho}(-\overline{q}_0,\overline{\boldsymbol{q}}=0),$$
$$\overline{\Pi}^{ab}_{00}(\overline{q}_0\to 0,\overline{\boldsymbol{q}}=2\pi(\overline{n},0,0)/L)$$
$$= \overline{\Pi}^{ab}_{00}(\overline{q}_0\to 0,\overline{\boldsymbol{q}}=2\pi(0,\overline{n},0)/L)$$
$$= \delta^{ab}\frac{g^2\overline{n}^2}{3L^2}\left\{-A^2_{\overline{n}}+\sum_{s=\pm}\frac{1}{8}\left(2-sA_{\overline{n}}\right)\left(1+sA_{\overline{n}}\right)^2\ln\left(1+sA_{\overline{n}}\right)^2\right\}, \quad (B.5)$$

and considering the relations

$$\left[(\delta_\perp+J)^\sigma_{\ \mu}(\delta_\perp-J)^\rho_{\ \nu}+(\delta_\perp-J)^\sigma_{\ \mu}(\delta_\perp+J)^\rho_{\ \nu}\right](g^\perp_{\sigma\rho}+g^{||}_{\sigma\rho}) = 4g^\perp_{\mu\nu},$$
$$\overline{\Pi}^{ab}_{\sigma\rho}(\overline{q}_0\to 0,\overline{\boldsymbol{q}}=0) = -\delta^{ab}(g^\perp_{\sigma\rho}+g^{||}_{\sigma\rho})g^2\mu_q^2/(6\pi^2), \quad (B.6)$$

we finally obtain

$$\Pi^{ab}_{\mu\nu}(q_0,q_0'\to 0,\boldsymbol{q}=\boldsymbol{q}'=0)$$
$$= 2\pi L^3\delta^{ab}\Big\{-g^{||}_{\mu\nu}\frac{g^2\mu_q^2}{6\pi^2}+g^\perp_{\mu\nu}\Big[-\frac{g^2\mu_q^2}{6\pi^2}-\frac{g^2\omega^2}{48\pi^2}\ln\left(1-4\mu_q^2/\omega^2\right)^2$$
$$-\frac{g^2\omega^2}{48\pi^2}\sum_{\overline{n}\neq 0}\left(-A^2_{\overline{n}}+\sum_{s=\pm}\frac{1}{8}\left(2-sA_{\overline{n}}\right)\left(1+sA_{\overline{n}}\right)^2\ln\left(1+sA_{\overline{n}}\right)^2\right)\Big]\Big\}, \quad (B.7)$$

where we have used the summation $\sum_{n\neq 0}(1/n^2)=\pi^2/3$. Using the definition (17) for the gluon mass, the extracted longitudinal and transverse masses are shown in (22) and (23).

## Data availability

Data will be made available on request.

## References

[1] F. Becattini, F. Piccinini, J. Rizzo, Phys. Rev. C 77 (2008) 024906, arXiv:0711.1253 [nucl-th].
[2] Y. Jiang, Z.-W. Lin, J. Liao, Phys. Rev. C 94 (2016) 044910, Erratum: Phys. Rev. C 95 (2017) 049904, arXiv:1602.06580 [hep-ph].
[3] W.-T. Deng, X.-G. Huang, Phys. Rev. C 93 (2016) 064907, arXiv:1603.06117 [nucl-th].
[4] V.V. Braguta, A.Y. Kotov, D.D. Kuznedelev, A.A. Roenko, Phys. Rev. D 103 (2021) 094515, arXiv:2102.05084 [hep-lat].
[5] M.N. Chernodub, Phys. Rev. D 103 (2021) 054027, arXiv:2012.04924 [hep-ph].
[6] Y. Jiang, Phys. Lett. B 853 (2024) 138655, arXiv:2312.06166 [hep-th].
[7] J.-H. Wang, S.-Q. Feng, Phys. Rev. D 109 (2024) 066019, arXiv:2403.01814 [hep-ph].
[8] X. Chen, L. Zhang, D. Li, D. Hou, M. Huang, J. High Energy Phys. 07 (2021) 132, arXiv:2010.14478 [hep-ph].
[9] A. Bzdak, S. Esumi, V. Koch, J. Liao, M. Stephanov, N. Xu, Phys. Rep. 853 (2020) 1, arXiv:1906.00936 [nucl-th].
[10] G.S. Bali, F. Bruckmann, G. Endrodi, Z. Fodor, S.D. Katz, A. Schafer, Phys. Rev. D 86 (2012) 071502, arXiv:1206.4205 [hep-lat].
[11] E.S. Fraga, A.J. Mizher, Phys. Rev. D 78 (2008) 025016, arXiv:0804.1452 [hep-ph].
[12] Z.-T. Liang, X.-N. Wang, Phys. Rev. Lett. 94 (2005) 102301, Erratum: Phys. Rev. Lett. 96 (2006) 039901, arXiv:nucl-th/0410079.
[13] Q. Wang, Nucl. Phys. A 967 (2017) 225, arXiv:1704.04022 [nucl-th].
[14] L. Adamczyk, et al., STAR, Nature 548 (2017) 62, arXiv:1701.06657 [nucl-ex].
[15] M.I. Abdulhamid, et al., STAR, Phys. Rev. C 108 (2023) 014910, arXiv:2305.08705 [nucl-ex].
[16] M.S. Abdallah, et al., STAR, Nature 614 (2023) 244, arXiv:2204.02302 [hep-ph].
[17] S. Acharya, et al., ALICE, Phys. Rev. Lett. 125 (2020) 012301, arXiv:1910.14408 [nucl-ex].
[18] S. Acharya, et al., ALICE, Phys. Rev. Lett. 131 (2023) 042303, arXiv:2204.10171 [nucl-ex].
[19] D. Kharzeev, A. Zhitnitsky, Nucl. Phys. A 797 (2007) 67, arXiv:0706.1026 [hep-ph].
[20] D.E. Kharzeev, D.T. Son, Phys. Rev. Lett. 106 (2011) 062301, arXiv:1010.0038 [hep-ph].
[21] Y. Jiang, X.-G. Huang, J. Liao, Phys. Rev. D 92 (2015) 071501, arXiv:1504.03201 [hep-ph].
[22] M. Wei, C.A. Islam, M. Huang, Phys. Rev. D 105 (2022) 054014, arXiv:2111.05192 [hep-ph].
[23] L. Dong, S. Lin, Eur. Phys. J. A 58 (2022) 176, arXiv:2112.07153 [hep-ph].
[24] S. Chen, J. Zhao, P. Zhuang, Phys. Rev. C 103 (2021) L031902, arXiv:2005.08473 [nucl-th].
[25] N.R.F. Braga, Y.F. Ferreira, Phys. Rev. D 108 (2023) 094017, arXiv:2309.11643 [hep-ph].
[26] E. Berti, F. White, A. Maniopoulou, M. Bruni, Mon. Not. R. Astron. Soc. 358 (2005) 923, arXiv:gr-qc/0405146.
[27] E.R. Most, A. Motornenko, J. Steinheimer, V. Dexheimer, M. Hanauske, L. Rezzolla, H. Stoecker, Phys. Rev. D 107 (2023) 043034, arXiv:2201.13150 [nucl-th].
[28] H. Berrehrah, E. Bratkovskaya, W. Cassing, P.B. Gossiaux, J. Aichelin, M. Bleicher, Phys. Rev. C 89 (2014) 054901, arXiv:1308.5148 [hep-ph].
[29] Y.J. Minghua Wei, M. Huang, Chin. Phys. C 46 (2022) 024102, arXiv:2011.10987 [hep-ph].
[30] A.C. Aguilar, D. Binosi, J. Papavassiliou, Front. Phys. (Beijing) 11 (2016) 111203, arXiv:1511.08361 [hep-ph].
[31] G. Huang, J. Zhao, P. Zhuang, Phys. Rev. D 108 (2023) L091503, arXiv:2307.02608 [hep-ph].
[32] L. Schubnikow, W. De Haas, Nature 126 (1930) 500.
[33] A. Ayala, L.A. Hernández, K. Raya, R. Zamora, Phys. Rev. D 103 (2021) 076021, Erratum: Phys. Rev. D 104 (2021) 039901, arXiv:2102.03476 [hep-ph].
[34] S. Chen, K. Fukushima, Y. Shimada, Phys. Rev. Lett. 129 (2022) 242002, arXiv:2207.12665 [hep-ph].
[35] J. Mäkelä, arXiv:2404.10364 [gr-qc], 2024.
[36] G.W. Gibbons, S.W. Hawking, Phys. Rev. D 15 (1977) 2752.
[37] Y. Jiang, J. Liao, Phys. Rev. Lett. 117 (2016) 192302, arXiv:1606.03808 [hep-ph].
[38] H.-L. Chen, X.-G. Huang, J. Liao, Lect. Notes Phys. 987 (2021) 349, arXiv:2108.00586 [hep-ph].
[39] G. Huang, P. Zhuang, Phys. Rev. D 104 (2021) 074001, arXiv:2107.00276 [hep-ph].
[40] M.L. Bellac, Thermal Field Theory, Cambridge Monographs on Mathematical Physics, Cambridge University Press, 2011.
[41] A. Ghatasheh, e-prints, arXiv:2305.17572, 2023, arXiv:2305.17572 [math.HO].
[42] R.A. Schneider, arXiv:hep-ph/0303104, 2003.
[43] J.O. Andersen, E. Braaten, M. Strickland, Phys. Rev. Lett. 83 (1999) 2139, arXiv:hep-ph/9902327.
[44] N. Haque, A. Bandyopadhyay, J.O. Andersen, M.G. Mustafa, M. Strickland, N. Su, J. High Energy Phys. 05 (2014) 027, arXiv:1402.6907 [hep-ph].
[45] J.O. Andersen, E. Braaten, M. Strickland, Phys. Rev. D 61 (2000) 014017, arXiv:hep-ph/9905337.
[46] J.O. Andersen, E. Braaten, M. Strickland, Phys. Rev. D 61 (2000) 074016, arXiv:hep-ph/9908323.
[47] A. Peshier, Phys. Rev. D 63 (2001) 105004, arXiv:hep-ph/0011250.
[48] A. Peshier, K. Schertler, M.H. Thoma, Ann. Phys. 266 (1998) 162, arXiv:hep-ph/9708434.
[49] Y. Jiang, H. Li, S.-s. Huang, W.-m. Sun, H.-s. Zong, J. Phys. G 37 (2010) 105004, arXiv:1007.1713 [hep-ph].
[50] V.N. Gribov, Nucl. Phys. B 139 (1978) 1.
[51] D. Zwanziger, Nucl. Phys. B 323 (1989) 513.
[52] C.S. Fischer, D. Nickel, R. Williams, Eur. Phys. J. C 60 (2009) 47, arXiv:0807.3486 [hep-ph].
[53] T. Holstein, R.E. Norton, P. Pincus, Phys. Rev. B 8 (1973) 2649.